 \documentstyle[12pt,aaspp4,psfig]{article}
 
 \def\gtsim{\lower.5ex\hbox{$\; \buildrel > \over \sim \;$}}
 \def\ltsim{\lower.5ex\hbox{$\; \buildrel < \over \sim \;$}}
 
 \def\esc2{ {\rm erg}\,{\rm s}^{-1}\,{\rm cm}^{-2}\,}
 
\begin{document}
 
 \tighten \title{The 2-10 keV XRB dipole and its cosmological
 implications} 
 \author{C. A. SCHARF } \affil{Space Telescope
Science Institute, 3700 San Martin Drive, Baltimore MD 21218, USA.
(scharf@stsci.edu)} 
 \author{ K.
 JAHODA} \affil{Laboratory for High Energy Astrophysics, NASA/Goddard Space Flight Center, Code 662, Greenbelt, MD
 20771. (keith@pcasrv2.gsfc.nasa.gov)} 
 \author{ M. TREYER} \affil{Laboratoire d'Astronomie Spatiale, Traverse du Sipho
n, B.P.8,
13 376 Marseille, France. (Marie.Treyer@astrsp-mrs.fr)}
 \author{O. LAHAV}\affil{Institute  of
 Astronomy, Madingley Road, Cambridge CB3 OHA, UK, and Racah Intitute
 of Physics, The Hebrew University, Jerusalem 91904, Israel. (lahav@ast.cam.ac.u
k)} 
 \author{E. BOLDT}
 \affil{Laboratory for High Energy Atsrophysics, NASA/Goddard Space Flight Center, Code 660, Greenbelt, MD 20771.
 (boldt@lheavx.gsfc.nasa.gov)}
 \author{T. PIRAN} \affil{Racah Institute of Physics, The Hebrew University,
 Jerusalem 91904, Israel. (tsvi@shemesh.fiz.huji.ac.il)}

\clearpage

 \begin{abstract}
 The hard X-ray ($>2$ keV) emission of the local and distant Universe as
 observed with the HEAO1-A2 experiment is reconsidered in the context of large
 scale cosmic structure. 
 
 Using all-sky X-ray samples of AGN and galaxy clusters we remove the dominant
 local X-ray flux from within a redshift of $ \sim 0.02$. We evaluate the
 dipolar and higher order harmonic structure in 4 X-ray colours. The estimated
 dipole anisotropy of the unresolved flux appears to be consistent with a
 combination of the Compton-Getting effect due to the Local Group motion
 (dipole amplitude $\Delta = 0.0042$) and remaining large scale structure
 ($0.0023 \ltsim \Delta \ltsim 0.0085$), in good agreement with the
 expectations of Cold Dark Matter models.

 The observed anisotropy does however also suggest a non-negligible Galactic
 contribution which is more complex than current, simple models of $>2 $keV
 Galactic X-ray emission. Comparison of the soft and hard colour maps with a
 harmonic analysis of the 1.5 keV ROSAT all-sky
 data qualitatively suggests that at least a third of the faint, unresolved 
 $\sim 18^{\circ}$  scale structure in the HEAO1-A2 data may be Galactic in 
 origin. However, the effect on measured flux dipoles is small ($\ltsim 3$\%).

 We derive an expression for dipole anisotropy and acceleration and demonstrate
how the dipole anisotropy of the distant X-ray frame can constrain the
amplitude of  bulk motions of the universe. From observed bulk motions over a
local $\sim 50$ h$^{-1}$ Mpc radius volume we determine $0.14 \ltsim
\Omega_0^{0.6}/b_x(0) \ltsim 0.59$, where $\Omega_0$ is the universal density
parameter and $b_x(0)$ is the present epoch bias parameter, defined as the
ratio of fluctuations in the X-ray source density and the mass density.
\end{abstract}


\clearpage
 
 \section{Introduction}
 
Establishing all the sources of the extragalactic flux dominating the X-ray 
sky remains a fundamental challenge. In the soft X-ray band ($0.5-2$ keV) some
70-80\% of the observed flux can be accounted for by extrapolation of those
objects resolved in deep fields (~\cite{has98}). Such objects are QSO's, AGN
and possibly narrow emission line galaxies. In the hard X-ray band ($\gtsim 2$
keV),  however,  the situation is less clear.   Deep surveys with ASCA,
reaching  $1 \times 10^{-13} \esc2$ in the 2-10 keV band now account for $\sim
30\%$ of the XRB (~\cite{cag98,ueda99}). Although recent results with Chandra
(e.g. ~\cite{mus00}) have now pushed this to $\sim 75$\%. In addition, the
spectral form of the  still unresolved flux does not fit with the spectra of
any single class of known objects, although more detailed models have had some
success (~\cite{lei92,com95,mad94}). The investigation of X-ray emission
associated with the local Universe has produced more tangible results.  The
autocorrelation function (~\cite{jah93}) of the unresolved X-ray flux and its
cross-correlation with other extragalactic catalogues
(~\cite{jah91,jah92,lah93,miy94,bar95,ref97,alm97}) provides useful information
on the volume emissivity of that fraction of X-ray emission correlated with
galaxies.  

For such studies the HEAO1-A2 all-sky survey continues to be the best hard band
all-sky data available. The integrated flux in this survey originates from
$z=0$ to a redshift of $z \simeq 4$; and might therefore be considered the most
complete survey of baryonic matter currently available at any wavelength. Less
than 2\% of the total extragalactic flux comes from sources identified in
existing all sky catalogs (~\cite{pic82,gro92}).   Since the majority of the
observed flux comes from high-$z$ we expect it to be highly isotropic, at least
to ${\cal O}(10^{-3})$, with possible deviations caused by anisotropies in the
population of nearby but unresolved sources. Indeed, the extragalactic hard
X-ray emission associated with foreground sources within $z\sim 0.02$ is highly
anisotropic, indicative of the pronounced structure in the mass distribution
traced by present-epoch AGNs. For example, the dipole to monopole ratio
observed in this population (~\cite{miy90,miy91,miy94}) is very large, $\sim
0.5$, while the monopole contribution to the total extragalactic emission is
$\sim 1$\%. The observed AGN dipole has also been demonstrated  in these
studies to be compatible with the direction and magnitude of the local group
velocity. This is a crucial observational `calibration' which supports the idea
that AGN X-ray sources do indeed trace the underlying gravitational mass
distribution responsible for peculiar motion.  

 Lahav, Piran \& Treyer (1997) (hereafter LPT97) have evaluated the expected
large-scale angular fluctuations in the XRB for a range of power spectra of
mass fluctuations (e.g. Cold Dark Matter (CDM) models) and X-ray evolution
scenarios. In a followup paper, and companion to this present work 
(~\cite{tre98}) we have investigated these large-scale
($10^{\circ}-180^{\circ}$) fluctuations. Using the HEAO1-A2 data the power 
spectrum of mass fluctuations can be probed on scales of $\sim 600$h$^{-1}$
Mpc  (where h is the present epoch Hubble  constant in units of $100$ km
s$^{-1}$ Mpc$^{-1}$). This data can also be used to constrain the fractal
correlation dimension ($D_2$) of structure on this scale (~\cite{pee93,wu99}) and
hence test the homogeneity of the Universe and the validity of the Cosmological
Principle. A value of $D_2=3$ to a precision of $10^{-4}$ is observed, strongly
supporting homogeneity on large ($\sim 600$h$^{-1}$Mpc) scales.

Work has also been done (~\cite{sha83,bol87,jah92,jah93}) to determine the
extragalactic X-ray flux dipole from the HEAO data. The motivation for this
being the determination of the expected Compton-Getting (CG) dipole due to our
motion with respect to the distant X-ray frame, which is expected to agree with
the direction and velocity inferred from the CMB dipole (~\cite{lin96}). The
amplitude of such a dipole constrains the distance of the frame which can be
considered to be at rest with respect to the CMB. 

However, as demonstrated in LPT97, for a typical observer in (for example) a
CDM universe the amplitude of the expected CG dipole and that due to emission
correlated with large-scale structure (LSS) are comparable.  This coupling of
the two dipole terms makes it hard to use the XRB dipole to confirm the CMB
motion dipole. Other investigations (~\cite{pli99}) of the XRB dipole using the
soft, 1.5keV ROSAT all-sky survey data seem to provide some confirmation of
this prediction. However, as we demonstrate below, data in this softer band is
strongly contaminated by Galactic emission compared to the hard, $>2$ keV HEAO
data.

In this present work we reconsider the HEAO1-A2 data and examine the X-ray
background dipole together with higher  order structures and investigate the
relative importance of the CG effect, LSS, Galactic emission and the
relationship of dipolar structure to  bulk motions.  In Section 2 we describe
the HEAO1-A2 all sky data, and a newly discovered small instrumental effect and
a prescription for its removal. In Section 3 we describe and apply spherical
harmonic analysis to the large angular scale structures in the data and
investigate the effects of removing known sources. In Section 4 we present
angular power spectra.  In Section 5 we measure the flux dipole of the
unresolved XRB and assess its significance through simple simulations. In
Section 6 we derive the full cosmological expression for dipole anisotropy and
its relationship to peculiar velocity. In Section 7 we discuss observations of
local bulk motions and apply these results to our X-ray dipole estimates to
obtain constraints on bias parameters. In Section 8 we summarise our results
and present conclusions.

  \section{The HEAO1-A2 data}

  We have taken the present data from the online archives at the High Energy 
 Astrophysics Science Archive Research Center (HEASARC) at the NASA/Goddard
 Space Flight Center \footnote{({\it
 ftp://legacy.gsfc.nasa.gov/heao1/data/a2/maps/heasarc\_med\_hed/ })}. In its
 raw state the data used here consists of the `Small Field of View' (SFOV)
 surface brightness in counts/sec per beam, stored in $0.25^{\circ}\times
 0.5^{\circ}$ (a total of $720\times  720$) pixels in a rectangular projection
 in  ecliptic coordinates. The intrinsic resolution size of independent data
 points is however $1.5^{\circ} \times 3^{\circ}$. In addition, for the purposes
 of estimating the instrumental background, we have utilized the equivalent
 `Large Field of View' (LFOV) data, which although stored in similar format
 ($720\times 720$ pixels) has an intrinsic resolution of $3^{\circ}\times
 3^{\circ}$. The all-sky survey is available in 4 overlapping energy bands or
 colours: Soft, Hard, Total and R15.  Allen et al (1994) presents effective area
 curves (detection efficiency as a function of energy) for these bands.   The
 Soft color consists primarily of photons detected in the first layer of the
 argon and xenon filled detectors and with pulse height less than 6 keV;  the
 Hard color consists of second layer photons and large pulse heights from the
 first layer and has very little response below 6 keV while the soft color has
 very little response above 8 keV. The R15 color was, throughout the mission,
 the unweighted sum of x-rays  detected in the first and second layer of the
 High-Energy-Detector (HED)-3  and the second layer of the
 Medium-Energy-Dectector. The weights which define these colours are chosen so
 that a source characterized by a photon index of -1.7 which produces 1 R15
 count/sec per beam will also produce 1 Total (or 1 Soft or 1 Hard) count/sec
 per beam. The R15 band is the most stable colour over the period of
 observations, the Total band has the highest signal-to-noise. The 4 bands have
 effective areas peaking at 3, 7, 6 and 10 keV for Soft, Hard, Total and R15
 respectively. All data was taken from the 6 month observation period beginning
 day 322 of 1977. To convert the raw surface brightness data into standard units
 we use a conversion factor of $2.2 \times 10^{-11}$ erg s$^{-1}$ cm$^{-2}$ (4.5
 deg$^2$)$^{-1}$ per count/sec. The beam area (4.5 deg$^2$) reflects the
 instrinsic $1.5^{\circ} \times 3.0^{\circ}$ resolution of the SFOV.

  In Figure 1 the raw HEAO1-A2 data is presented in a Hammer-Aitoff, Galactic
  projection. The data consists of one complete scan of the sky using
  the Total band (~\cite{all94}).

  The darkest pixels are associated with individual bright sources and diffuse
  Galactic emission.  Only $\sim 100$ of the brightest sources at high galactic
  latitudes can be identified (~\cite{pic82}) identifies 17 galactic and 68
  extragalactic sources).  In all of the following we restrict ourselves to
  $|b| > 20 ^{\circ}$ where complete catalogs are available.

  The data is first corrected for the locally estimated instrumental
  background. This is achieved by using both the LFOV and SFOV data to solve
  the following two equations at each pixel for $I_{background}$:
  $I_{sfov}=I_{background}+I_{sky}$ and $I_{lfov}=I_{background}+2.26 I_{sky}$,
  where the factor of 2.26 is the ratio of area solid-angle products for the
  LFOV and SFOV. To reduce noise in the background estimate we then correct the
  data by subtracting the mean background estimated in strips of constant
  ecliptic longitude. Prior to the background calculation we remove  sources
  and mask Galactic regions as described below.

    In the course of examining the data we observed a systematic change in the
  measured flux as a function of observation date. Since the data was taken in
  great circles through the ecliptic pole, data separated by $180^{\circ}$ in
  ecliptic longitude correspond to approximately the same epoch of observation.
  Taking the mean  flux at fixed longitude of the source removed/masked data 
 (see below),
  wrapped by $180^{\circ}$ we discovered a clear linear trend in the observed
  counts. In Figure 2 we plot the observed effect in the Total band data. The
  least squares linear models for all 4 HEAO colours are shown in Figure 3. The
  slopes of the trends (in units of counts/sec, vs longitudinal pixel index) are
  $-0.000716, -0.000194, -0.000357, -0.000265$ for the Soft, Hard, Total and R15
  colours respectively, with maximum and minimum at longitudinal pixel indices
  69 and 429 respectively.

  The time dependent term therefore ranges from $0-7\%$ of the mean intensity
  for the Soft band, and less for the other colours.  In all subsequent analysis
  the HEAO data has been corrected by subtracting the least squares trend. The
  resulting change on all quantities discussed is small and always less than
  10\%. 

  To improve computational efficiency in all the following analysis we have
  re-binned the data (following background, and systematics correction, and
  source removal as described below) into $3^{\circ} \times 6^{\circ}$ pixels
  in ecliptic coordinates (smaller resolution pixels are more strongly
  correlated due to the instrument beam)

  \section {Large scale structures in the HEAO data}

  Unlike many all-sky catalogues (e.g. IRAS, optical surveys etc.) the ability
  to unambiguously separate foreground (Galactic/local emission) from
  background (extragalactic) information is limited in the HEAO X-ray data. The
  total number of resolved foreground and background sources is small ($\sim
  100$) and a detailed model of possible large scale Galactic emission is hard
  to determine. The Galactic $2-60$ keV emission model of Iwan et al (1982)
  predicts that some 2\% of the observed emission at the galactic poles in the
  A2 band is of Galactic origin.  The largest contribution at latitudes $b \ge
  20^{\circ}$ is 5\%.  (The Iwan model predicts galactic contributions of up to
  10\% at low latitudes, although there is certainly another, more centrally
  concentrated component as well (~\cite{wor82,war85,val98}).  More recently,
  studies in the soft bands ($<0.75$ keV) by ROSAT (~\cite{sno97}) indicate
  that, at these lower energies, the picture is more complicated, with
  structure at all scales. Whether this soft emission distribution is a good
  indicator of the much harder $>2$ keV emission is unclear;  the galactic
  contribution in this soft band is almost certainly larger than that  above 2
  keV, and is potentially more variable as well.  In this present work we
  attempt to at least qualitatively evaluate the likely foreground vs.
  background contributions to further constrain our estimates of the
  extragalactic flux anisotropy.  In order to evaluate the structure in the map
  of Figure 1 we use spherical harmonic analysis to filter the high (noisy)
  frequencies from the map and to reconstruct the flux variations (e.g.
  Scharf et al 1992).  Briefly, the map is expanded into the orthonormal set of
  spherical harmonics ($Y_{lm}(\theta, \phi)$) by determining the harmonic
  coefficients as a sum over the flux cells:  

  \begin{equation}  
 a_{lm}= \sum_{i} I_{i} \Delta \omega_{i} Y_{lm}^{*}({\hat \omega_{i}})\;\;\;, 
  \end{equation}  

  where $I_{i}$ is the mean surface brightness in a cell, $\Delta \omega_{i}$ is
  the cell area in direction ${\hat \omega_{i}}$.  The surface brightness at any
  point can then be reconstructed using only lower order harmonics, where the
  resolution is determined by the highest harmonic and goes as $\sim
  \pi /l_{max}$. In our definitions below the monopole $M_{x}$ is defined as
  $4\pi \bar I$ thus $a_{00} \equiv \sqrt{4\pi} {\bar I} = M_x/ \sqrt{4\pi} $.

  \subsection{Higher order anisotropies}

In Figure 4 a harmonic reconstruction of the raw (Total band) HEAO data is
shown to a resolution of $ \sim 18^{\circ}$ ($l_{max}=10$). The data is clearly
dominated by emission associated with the Galaxy (either resolved or unresolved
sources). As a first step towards removing this foreground we construct a
`mask' using the list of resolved and identified  Galactic X-ray sources
(~\cite{pic82}) and a $|b|<22^{\circ}$ Galactic Plane mask. Regions of sizes
varying from $\sim 8^{\circ}$ diameter to $12^{\circ}$ diameter are excised
around resolved sources, larger regions are removed around the Large and Small
Magellanic Clouds (LMC, SMC). A total of $\sim 38$\% of the sky area is removed
by this mask. The results are dramatic, in Figure 5 the harmonic reconstruction
(to $l_{max}=10$) of the data is shown in all four bands, with the above
Galactic masking applied. Those regions excised have been filled with the new
mean flux as a first order correction (see discussion of dipole estimation
below). As described in Section 2, the count/sec in each colour are weighted to
be equivalent for emission with a photon index of -1.7. The observed
differences in the structure between (for example) the Soft and Hard bands can
then be directly interpreted as spatial differences in the mean spectral index
of X-ray emission (modulo variations in the signal-to-noise).

  Two strong flux enhancements are apparent. The uppermost (at $b\gtsim
  80^{\circ}$ and spanning the Northern Galactic cap) we associate with the 
  Virgo and Coma clusters. The lower peak (at $l\sim 315^{\circ}$ and $b\sim
  30^{\circ}$) is close to the Centaurus/Great Attractor region (e.g. Scharf et
  al 1992, Webster et al 1998). However, we note that it is also closer to the
  Galactic Plane and may not be free of Galactic contamination.

  Next we make use of the Pincinotti (1982) catalogue of extragalactic sources
  (supplemented with the clusters of Edge et al (1990) with $z<0.003$) and
  excise these regions (in $\sim 8^{\circ}$ diameter cuts) in addition to the
  above Galactic masking.  This results in a dataset with all sources removed
  down to a flux limit of $3 \times 10^{-11}$ ergs s$^{-1}$ cm$^{-2}$ (c.f.
  Treyer et al 1998). The median redshift of sources in the extragalactic
  Piccinotti sample is known to be $z\sim 0.02$. In a further effort to remove
  all significant X-ray sources from within the local     volume, we have made
  use of the HEAO A1 catalogue of source detections (Ron Remillard, private
  communication). Using this data we have removed all sources to a slightly
  lower flux limit of $2\times 10^{-11}$ ergs s$^{-1}$ cm$^{-2}$  (2-10 keV).
  We determine this flux limit value from both our chosen limiting count rate
  in the A1 catalogue (0.006 ct/s) and by normalising with respect to the
  extrapolated Piccinotti LogN-LogS. This yields a final monopole value
  corresponding to $M_{x} \simeq 6.2 \times 10^{42}$ erg s $^{-1}$ Mpc$^{-2}$. 
  The final unmasked sky area is 48\%, in Figure 6 the corresponding harmonic
  reconstructions (to $l_{max}=10$) are shown.

  From the X-ray luminosity functions of Grossan (1992) and Boyle et al.
  (1998), we estimate that the mean luminosity of the local X-ray source
  population is $L_\star \sim 5 \times 10^{42}$ erg s$^{-1}$, assuming a lower 
  luminosity cutoff of $10^{42}$ erg s$^{-1}$ and a present day Hubble constant
  H$_0$=100 kms$^{-1}$Mpc$^{-1}$. As the luminosity function is dominated by
  faint objects and and the low-luminosity cutoff is somewhat arbitrary,
  $L_\star$ is likely to be a lower limit.    Removing sources brighter than $2
  \times 10^{-11}$ erg s$^{-1}$ cm$^{-2}$ corresponds to removing all sources
  brighter than $L_\star$ out to $z \sim 0.015$, and all sources brighter than
  $10 \times L_\star$ out to $z \sim 0.05$. Therefore most X-ray sources with
  $z > 0.015$ are still unresolved and contribute to the measured dipole (see
  below).

  The contour levels in Figure 6 are approximately a factor $\sim 3$ more
  exaggerated relative to the mean than in Figure 5.  Unlike the extragalactic
  sky dominated by resolved sources, significant differences in structure is
  now seen between the Soft and Hard colours. Notably, in the Soft band two
  strong patches are seen at $(\sim 10^{\circ}, \sim -28^{\circ})$ and $(\sim
  80^{\circ}, \sim -40^{\circ})$, and a significant structure at $l\simeq
  180^{\circ}$, extending over $15^{\circ}\ltsim b \ltsim 80^{\circ}$. In the
  Hard band the most significant features are seen at $(\sim 200^{\circ}, \sim
  -40^{\circ})$ and $(\sim 280^{\circ}, \sim 60^{\circ})$. Possible
  identifications with known structures are given in the Figure 6 caption.

In Figure 7 we plot the harmonic reconstructions of `noise' skies for both the
Galactic and full extragalactic mask cases in Figures 5 \& 6, to provide a
qualitative guide to the significance of features. From these figures we can
assess the signal-to-noise of structures seen in Figures 5 \& 6 (Total band).
The most significant individual features are typically a factor 2-3 higher in
counts/sec when compared to the equivalent peaks in the pure noise levels. It
is apparent that there are at least two likely real, positive, features in
Figure 5 and possibly four  in Figure 6.

  In an attempt to assess the likely amplitude and morphology of Galactic
  contamination in the HEAO1-A2 data we have analysed the all-sky X-ray data
  from the ROSAT All Sky Survey (RASS) using the maps of Snowden et al (1997,
  1995). In Figure 8 we present a harmonic reconstruction of the 1.5 keV RASS
  data, subject to the same Galactic and extragalactic source mask as the
  HEAO1-A2 data as shown in Figure 6. The RASS data was used in a $480\times
  240$ pixel Hammer-Aitoff Galactic projection as available from the online
  archives, no further processing was made, other than filling empty pixels
  with the mean flux after applying the HEAO1-A2 Galactic and extragalactic
  source masks. As  discussed in Snowden et al (1995) the 1.5 keV band data
  spans the energy range $\sim 0.7 - 2$ keV, with essentially no counts above 2
  keV. The HEAO1-A2 Hard band is therefore the only colour with no response in
  this energy range. The response in the other bands is also modest at 2 keV,
  the effective area is down from its maximum by factors of $\sim 3$, 10, and
  10 for the Soft, Total, and R15 colours.  The three most prominent features
  in the RASS reconstruction are concident with the features identified in
  Snowden et al (1995) as the Northern Spur/Loop I ($l\sim 20^{\circ}, b\sim
  40^{\circ}$), and the northern and southern parts of a Galactic Bulge
  emission ($l\sim340^{\circ}, b\sim 30^{\circ}$,  $l\sim 0^{\circ}, b\sim
  -30^{\circ}$).

  The location of the two Bulge structures is particularly worrying for our
  interpretation of the structures in Figure 6. In particular the southern
  structure seen most strongly in the Soft band at $l\sim 10^{\circ}, b\sim
  -30^{\circ}$ is {\em not} present in the Hard band data and might indeed be
  correlated with the RASS southern Bulge feature. Snowden et al (1997) have
  further investigated the RASS maps and propose that a soft Galactic bulge
  component is well modelled by a cylindrical volume (with exponential density
  fall-off away from the Galactic plane) of $\sim 10^{6.6} K$ gas. As discussed
  above, this does not preclude Galactic bulge emission above 2 keV, but for a
  thermal spectrum very little emission would be expected. The southern HEAO
  'Bulge' feature appears to fall into this category. The HEAO structure at
  $l\sim 325^{\circ}, b\sim 30^{\circ}$ is however present in all 4 colours, 
  which suggests that while it could be contaminated by the northern Bulge
  feature, it is more likely to have an extragalactic origin. Therefore, we
  conservatively estimate that at least 1 out of 3  of the dominant faint 
 (soft), unresolved structures in the HEAO1-A2 data may be Galactic in origin.
  This is consistent with the detailed investigation of the RASS 1.5 keV dipole
  made by Plionis \& Georgantopoulos (1999) who estimated that $20-30$\% of the
 total background is Galactic in origin in that data.

  Without much more rigorous cross-comparison any correlations of the RASS
  with  HEAO1-A2 should be made with caution, since the intrinsic spatial
  resolution of the RASS ($\sim 2^{\circ}$) is significantly different to that
  of the HEAO1-A2. In Section 5 below we investigate the effect on the flux
  dipole  of Galactic emission as described by the Iwan et al (1982) model, as
  well as the effect of removing the putative `Southern Bulge' feature.

  \section{Angular power spectra}

  In an associated paper (\cite{tre98}) we have investigated the angular
  fluctuations of the XRB in the context of models of large scale structure
  (as  formulated by LPT97).
  
 Following the formalism used in those papers the total predicted angular power
  (taken as an ensemble average) is a combination of a large-scale structure
  component (signal) and a component due to the discreteness of sources
  (shot-noise):

  \begin{equation}
\left<|a_l^m|^2\right>_{model}=\left<|a_l^m|^2\right>_{LSS}+
\left<|a_l^m|^2\right>_{SN}
\end{equation}

This is then convolved with the foreground mask to produce the observed
angular power spectrum:

\begin{equation}
\left<|c_l^m|^2\right>=\sum_{l'm'}|W_{ll'}^{mm'}|^2\left<|a_{l'}^{m'}|^2\right>
\end{equation}

Finally we normalize over the monopole, and for total signal use the notation:

\begin{equation}
C_l=\frac{(\left<|c_l^m|^2\right>_{LSS}+
\left<|c_l^m|^2\right>_{SN})^{1/2}}{\left<|a_{00}|^2\right>^{1/2}}\;\;\;,
\end{equation}

(see Section 3).

  Figure 9 plots the normalized angular power spectra (i.e. relative to the
  monopole, \cite{tre98}, Eqn 14) of the HEAO data for the
  complete extragalactic plus Galactic mask in all 4 colours. The expected CG
  effect (see \S 5 below) has also been subtracted from the data. (Note that in
  Treyer et al (1998) fewer sources were removed, corresponding to a slightly
  higher flux limit of $3\times 10^{-11}$ erg s$^{-1}$ cm$^{-2}$ (2-10 keV)).
  The Total band spectra used in Treyer et al (1998) are consistent with those
  shown here, despite some differences in data processing and source removal.
  The solid curve in Figure 9 is the predicted Total band angular spectra for a
  low density CDM model with pure X-ray source density evolution  and constant,
  linear biasing (see \S 6 below and Treyer et al (1998)), and including the
  predicted noise due to unresolved sources (LPT97).

   It is
  reassuring that the 4 spectra are in generally close agreement (despite
  different noise characteristics). We do however note that in 
  Figure 9, the Soft band spectra (possibly subject to the most Galactic
  contamination) shows a strongly discrepant quadrupole ($l=4$) term.

  In Figure 10 we also plot the mean spectra obtained from `noise' simulations.
 As described in Section 5 below these consist of sky fluxes randomly drawn
 from the real data and masked as per the real data. The complete sky mask is
 applied both to the data from which fluxes are drawn and to the simulations. 
 The noise estimated from simulations is in agreement with  the analytic
 estimate given in LPT97  and Treyer et al. (1998).

  It is apparent that the harmonic shot noise level varies between the HEAO
  colours. As expected the Total band has the lowest noise level, followed by
  the R15 band and the highest noise in the Soft and Hard bands, at very
  similar levels. Comparing these spectra with those in Figure 9 we can
  see that by $l\sim 5$ in the real data we are approaching the shot noise
  level and by $l\sim 10$ we have practically reached the shot noise level.

 \section{The observed flux dipole}

 The dipole flux anisotropy is described by the harmonic coefficients $a_{1m}$,
 and can also be written or derived as a simple vector ${\bf D}_{x}$, where;
 
 \begin{equation}
 {\bf D}_x=\sum_{i} I_i \Delta\omega_i \hat{\bf r}_i
 \end{equation}
 
  We can of course parameterize this differently. 	
For example, the CG effect
due to  our motion with respect to a radiation field (c.f. CMB dipole etc) is a
purely dipolar effect of the form $I'=I(1+\Delta \cos \theta)$ where $I$ is
surface brightness, and $\theta$ is measured from the direction of motion, in
which  case $|{\bf D}_{x}|/M_{x}=\Delta /3$ (note that $M_x=4\pi \bar I$). The
factor  $\Delta$ for the 2-10 keV  X-ray band CG effect is:
 \begin{equation}
 \Delta = \frac{3v}{c} [1+(\alpha/3)] \;, 
 \end{equation}

 where the energy spectral index is $\alpha=0.4$ (~\cite{bol87}). Assuming the
 Solar velocity with respect to the CMB (\cite{lin96}), we determine $\Delta=4.2
 \times 10^{-3}$ for the X-ray CG effect. Here we choose to present the dipole
 amplitude as $\Delta$ using the above conversion from ${\bf D}_{x}$.

 If an all-sky, foreground removed, X-ray surface brightness map existed the
 flux dipole could be immediately obtained as a vector sum over the flux cells.
 However, as demonstrated above, removing the foreground involves removing
 information about the background too.  The first order correction to the
 removal (masking) of regions when the harmonic decomposition is performed is
 to fill those regions uniformly with the mean density/flux of the unmasked
 regions. This will not, however, remove cross-talk between the true (full sky,
 orthonormal) harmonic coefficients (including the dipole) (e.g. Scharf et al
 (1992)). One solution is to attempt to reconstruct the full sky harmonics by
 inversion of the coefficient matrix, suitably controlled by (for example) a
 Wiener filter (e.g. Lahav et al (1994)). While this is a very powerful method
 it does require a model of the expected harmonic power and a full knowledge of
 the noise matrix. Additionally, such reconstruction is severely limited by the
 amount of masking (for a realistic harmonic spectrum of large scale
 structure), in the case of the HEAO data this is large; if all resolved
 Galactic and extragalactic sources and the Galactic plane are masked $\sim
 52$\% of the sky is removed.

 In this present work we use two different methods of dipole estimation. 
 First, we perform the vector sum over all flux cells, with a first order
 correction of filling masked cells with the mean flux over the unmasked
 regions (the spherical harmonic approach, hereafter Method 1). Second, we
 perform a least squares fit of a $cos(\theta)$ dependence dipole to the data
 and determine the best fit $(l,b)$ and $\Delta$ (hereafter Method 2). This
 latter method does {\em not} make use of the masked region, and does not
 suffer from cross-talk, but it does assume the specific form of the dipole 
 anisotropy, and (unrealistically) that the residuals are negligible. In the
 following we apply it only to the Total band data, since this data has the
 best signal-to-noise, and evaluate both methods using Monte-Carlo simulations
 below.

 The results of the various dipole estimations for the HEAO data are presented
 in Tables 1 and 2 for Methods 1 and 2 respectively.

 As a comparison with previous works on the dipole moment of X-ray AGN
(~\cite{miy90,miy91,miy94t}) we subtract  the Total band, source removed
extragalactic dipole (row 7, Table 1) from the full extragalactic Total band
dipole (row 3, Table 1). The resulting vector (which is equivalent to the
dipole vector of the removed extragalactic sources)  has $\Delta \simeq 8.2
\times 10^{-3}$ and points at $(316^{\circ},53^{\circ})$. Miyaji (1994) found
that the flux dipole of resolved AGN (which has different shot noise) in the
HEAO1 survey to $z\ltsim 0.015$ has a direction $(293^{\circ},33^{\circ})\pm
20^{\circ}$, consistent with this difference vector. In their analysis of the
RASS 1.5 keV XRB dipole Plionis \& Georgantopoulos (1999) determine a best
dipole direction of $(288^{\circ}, 25^{\circ}$) and amplitude $\Delta=0.051$
(converting their $D_x/M_x$). While the direction is in general agreement with
our results their amplitude is more than a factor of $\sim10$ higher compared
to our results in Tables 1 \& 2. This discrepancy is likely due to the 
increased difficulty of diffuse foreground removal in the soft ROSAT band ($<2$
keV) and an increased soft-band contribution from galaxy clusters and groups.
Indeed, Plionis \& Georgantopoulos (1999) estimate that Virgo contributes as
 much as 20\% to the RASS dipole amplitude.

 As a demonstration of the effect of applying a correction assuming the Iwan et
al (1982) Galactic emission model the 3rd block of Table 1 presents the results
of removing  a Iwan Galactic emission model (normalized to 3\% of the monopole)
from the distant flux dataset. The effect is relatively modest and
systematically decreases the dipole amplitude $\Delta$, and moves the observed
dipole direction away from the Galactic center. We also assess the effect of
removing the expected CG effect from the data (block 4, Table 1). The dipole is
significantly altered; $\Delta$ drops by $15-20$\% and the direction changes by
as much as $\sim 50^{\circ}$, towards the Galactic center. Combined with a 3\%
Iwan et al. correction (block 5 Table 1) $\Delta$ is further reduced. We also
measure the variation in $\Delta$ (and direction, see Figure 13) as a function
of increasing Galactic correction (Figure 11), for cases with and without
removal of the expected CG effect (lower and upper curves respectively). In the
case in which the CG effect is removed (after the Galactic correction) the
amplitude of the Galactic normalization is the most pronounced, reducing
$\Delta$ by a factor of 4 between a 1\% and 13\% correction.

 In Table 2 the equivalent results from the Method 2 dipole measurements
 (using the Total band) are presented. While the directions agree fairly well
 with those of the Method 1 dipole estimates, the amplitudes appear to be
 systematically larger by factors $\sim 2.5-3$ (see below). 

 Finally, we test the effect of removing a $30^{\circ}\times 30^{\circ}$ patch
 of sky around the putative `Southern Galactic Bulge' region, centred on
 $l=0^{\circ}, b=-25^{\circ}$. As done above this region is filled with the
 mean flux. The effect on both Method 1 and 2 dipole estimates is small,
 reducing $\Delta$ by $3-5$\% and altering the direction by $\delta \theta 
 \sim 10^{\circ}$. 

 In order to test the significance of the dipole measurements and the ability
 of the two methods in recovering a genuine dipole signal we use simple
 Monte-Carlo simulations.  Taking the fluxes of a dataset with the Galactic and
 extragalactic mask applied  we resample the flux distribution and construct a
 random sky map (equivalent to assuming no correlation between flux cells)
 which is subject to the same sky incompleteness (in this case the complete
 mask). The results of harmonic analyses on the simulations over  all bands and
 to $l=20$ have been shown previously in Figure 10. Here we concentrate on the
 dipolar measurements.

 The mean dipole amplitudes over 10 realizations,  are shown in Table 3. It is
clear that the `noise sky' dipoles are significantly smaller than the dipoles
seen in the real data, indicating the presence of genuine correlated structure.
The bottom two rows of Table 3 show the result of adding a realistic CG effect
to the simulated data. It is encouraging that the estimated amplitudes are
close to the input value ($\Delta = 4.2\times 10^{-3}$), and the dipole
directions are in general agreement, but we note that the two methods appear to
differ in a systematic way, with the Method 1 estimate being consistently
smaller by a factor $1.5-1.8$. This is not unexpected. As discussed in Treyer
et al (1998) and Scharf et al (1992) (and references therein) an incomplete sky
creates cross-talk between the harmonic coefficients. In the case of the mask
used here the net result is to systematically lower the observed amplitude of
the Method 1 dipole. The Method 2 dipole estimate does not suffer from such an
effect, although it is a less general estimate of dipole anisotropy. The
difference seen in the simulations accounts for at least 50\% of the Method 1/2
$\Delta$ discrepancy seen in the HEAO1-A2 data. More detailed simulations would
be needed (including realistic fluctuations instead of pure noise) to determine
the precise difference expected. A full treatment of the significance of the
observed dipole is beyond the scope of this present work.

However, on the basis of the simulations (Table 3) and the results of Tables 1
and 2, we estimate that our observed dipoles are significant at greater than a
$\sim 2-3 \sigma$ level, and have a typical direction error of $\sim
30^{\circ}$.

\section{The flux dipole and bulk motions}


The X-ray flux dipole observed at frequency $\nu_0$
is defined as: 
\begin{equation}
{\bf D}_x \equiv \sum_{i} f_{i}(\nu_{0}) \hat{{\bf r}_{i}}\;,
\end{equation}
where the sum is over all directions in the sky, and $f_{i}(\nu_{0})$
is the integrated X-ray flux in the direction $\hat{{\bf r}_{i}}$.

Following the formalism given in LPT97 and assuming 
linear, epoch-dependent biasing $b_x(z)$, such that fluctuations in
X-ray sources and mass are related by $\delta_x(r_c,\hat{\bf r})=b_x\delta(
r_c,\hat{\bf r})$, then:

\begin{equation}
{\bf D}_x= \int\int  \phi(L_{\nu },z)  \frac {L_{\nu } (1+z)}{4\pi
r_{L}^{2}} [1+b_{x}(z) \delta(r_c, \hat{\bf r})] \hat{\bf r}\ dV_{c}
dL_{\nu }\; ,
\end{equation} 

where $\phi$ is the radial probability of a source with luminosity
$L_{\nu}$ at redshift $z$ and $\delta$ is the mass density contrast.

If the number density of the X-ray sources
evolves as $(1+z)^d$, their luminosity as $(1+z)^e$,
and $L_{\nu }\propto \nu^{-\alpha}$, then we can define 
$q=d+e-\alpha+1$ and the X-ray volume emissivity as: 
\begin{equation} 
\rho_x(z)\equiv \int  L_{\nu } \phi(L_{\nu },z)
(1+z)\; dL_{\nu } = \rho_{x}(0)(1+z)^q ~.
\end{equation}
The dipole can then be written in the form of a ``dipolar Olbers' integral'':
\begin{equation} 
{\bf D}_x=\frac{1}{4\pi}\int \frac{\rho_x(z) b_x(z)\delta(r_c,\hat{\bf r})}
{r_c^2 (1+z)^2} \hat{\bf r}\; dV_c \;
\end{equation}
Recall that in an Einstein-de Sitter Universe ($\Omega_0=1$) 
$dV_{c}=r_c^{2} dr_{c}  d\omega$ and $r_{c}= 2r_H[1-(1+z)^{-1/2}]$,
where $r_H= c/H_0$ is the Hubble radius. 

As we do not have a model for $\delta({\bf r})$ 
in our neighbourhood, we can only make statistical
predictions using a model for the
power-spectrum $P(k)$ (LPT97, Treyer et al 1998).
Of course, what we observe is a single realization and this
one realization may not be well represented by the rms value. 
The rms dipole ($l=1$) can be expressed as (see Eq. 7 in Treyer et al 1998):
\begin{equation}
\langle |a_{1m}|^2 \rangle_{LSS}  =  
{(r_H~ \rho_x(0))^2 \over (2 \pi)^3}
\int  k^2 {\bar P}(k) |\Psi_1(k)|^2 {\rm d}k,
\end{equation}
where the window function $\Psi_1$ contains the various model
parameters:
\begin{equation}
\Psi_1(k) = \int_{z_{min}}^{z_{max}} \sigma_8 b_x(z)
(1+z)^{q - 9/2}  j_1(k r_c) W_{cut}(z) {\rm d}z~,
\end{equation}
and the function $W_{cut}(z)$ accounts for the removal of sources 
brighter than a given flux cutoff $S_{cut}$. $\sigma_8$ is the usual
rms mass fluctuation in an $8$h$^{-1}$Mpc radius sphere. As in Figure 8,
 we use a fiducial model  assuming low density CDM, pure density evolution 
 with $q=4.6$ and $z_{max}=1.3$ (based on  Hasinger 1998), and constant biasing.

Figure 12 shows the growth of the rms flux dipole as a function of the  outer
radius cutoff $z_{max}$ for our fiducial cosmological model and 3 scenarios for
$W_{cut}(z)$. The figure shows, first, that a flux cutoff  of $2 \times
10^{-11}$ erg s$^{-1}$cm$^{-2}$,  as used in the present data analysis, is very
similar to   removing all sources within $z<0.01-0.015$; and secondly, that the
rms dipole converges very rapidly, so that most of it originates from $z< 0.05
- 0.1$. Consequently there is very little signal due to structure lying further
out, and in the presence of noise we will have effectively no information from
$z>0.1$, at least in the rms sense. As the growth of the dipole depends on the
power spectrum, $P(k)$, in models with more large scale power the convergence
with $z$ will be slower.  Note that this convergence is not due to an ``Olbers'
effect''. The total intensity ${\bar I}$ of the XRB keeps increasing to much
higher redshift than do the fluctuations: ${\bar I} \propto
\int_0^{z_{max}}(1+z)^{q-7/2}{\rm d}z~$ while $\Psi_1(k)  \propto 
\int_0^{z_{max}} (1+z)^{q-9/2}  j_1(k r_c) {\rm d}z$ (to first approximation). 
Unlike the monopole, the high redshift fluctuations (dipole and higher
harmonics) are effectively washed out by angular averaging over the sky
(governed by the Bessel function dependency).  

Therefore, we can only use the XRB flux dipole to constrain large scale
structures out to 150-300 h$^{-1}$ Mpc. Coupled with our estimate that the
bright sources we remove are distributed to a distance of $\sim 60$
h$^{-1}$Mpc, we should be able to compare our XRB dipole measurement  with
direct measurements of bulk flows over a similar volume.

In linear perturbation theory
the peculiar motion at any point in space is directly proportional to the
gravitational acceleration, we can therefore write (assuming all motion was
zero a Hubble time ago (~\cite{pee80}): 
 \begin{equation} {\bf v}\simeq
 \frac{2}{3} \Omega_{0}^{-0.4} {\bf g} H_{0}^{-1} \;, 
 \end{equation} 
where $\Omega_{0}$ 
is the density parameter. 
The gravitational acceleration ${\bf g}$ in Newtonian gravity is:
\begin{equation} 
{\bf g}= G \rho(0)\int  \frac{\delta({\bf r})}{r^2} \hat{\bf r} \; dV \;,
\end{equation}
($\rho(0)$ is the present-epoch mean mass density).
We note that this expression only holds in on small scales, 
by choosing locally Minkowski coordinates (Peebles 1993, p.268).
We can therefore approximate Eqn. 10 above for the flux dipole 
at low redshift to be:
\begin{equation} 
{\bf D}_x=\frac{1}{4\pi}  {\rho_x(0) b_x(0)} {\bf g}/G 
\end{equation}

Therefore at low redshift the flux of a source follows an inverse square law
and if light traces mass in a spatially invariant and linear way then any
anisotropies seen in the X-ray data  reflect the local gravitational
acceleration (assuming linear theory).  

We again note that the CG effect produces a dipole pattern on the sky of the
form (see Eqn 6):

\begin{equation}
\frac{\Delta I}{\bar{I}}=(3+\alpha)\frac{V_{obs}}{c} cos(\theta)\;\;\; .
\end{equation}
Consequently the observed dipole will always be a coupling of the LSS and CG
dipole anisotropies.

The well known direct linear-theory relationship between the peculiar 
velocity and the absolute flux dipole is therefore (from Eqns 13, 14 \& 15):
\begin{equation} {\bf v}=
\frac{\Omega_{0}^{0.6}}{b_{x}(0)} \frac{H_{0}}{{\rho_{x}(0)}} {\bf D}_{x} \;
\end{equation} 
(c.f. Boldt 1987, Lynden-Bell et al 1989, Miyaji 1994). 
To express the linear velocity in terms of the LSS dipole anisotropy $\Delta$ we
recall the Olbers integral for $\bar{I}$ (Treyer et al 1998, Eqn 11):
\begin{equation}
\bar{I}=\frac{\rho_{x}(0) r_H Q(q,z_{max})}{4\pi}
\end{equation}
where the Hubble radius $r_H=c/H_0$ and
$Q(q,z_{max})=((1+z_{max})^{q-2.5}-1)/(q-2.5)$. Since the monopole $M_x=4\pi \bar{I}$
and $D_x/M_x=\Delta/3$, for the fiducial values of $q$ and $z_{max}$ used here we
arrive at:
\begin{equation}
|{\bf v}| = 2.2 \times 10^5  \Delta \frac{\Omega^{0.6}_0}{b_x(0)}\;\; km s^{-1}\;\;.
\end{equation}

\section{Comparison with observed bulk motions}

As discussed above, we estimate that the bright sources we remove from the
HEAO data are distributed to a distance $\sim 60 $h$^{-1}$Mpc, we can therefore
compare our XRB dipole measurement  with direct measurements of bulk flows over
this scale.

Several studies provide generally consistent estimates of the bulk flow of a
$\sim 50$h$^{-1}$Mpc radius sphere; $305-370 (\pm 110)$ kms$^{-1}$  (MIII
catalogue, POTENT, Dekel et al 1999), $\sim 300$ kms$^{-1}$ (SFI data, Dale et
al 1999, Giovanelli et al 1997), $\sim 250$kms$^{-1}$ (SNIa data, Riess et al
1995). The directions of these flows are summarised in Figure 13. Using a crude
mean of these numbers we estimate $v_{60} \simeq 300 \pm 100$ km s$^{-1}$. The
range of dipoles measured in the Total band (after removal of the dominant
X-ray sources from within $\sim 60$ h$^{-1}$ Mpc) is $0.0023 \leq \Delta \leq
0.0095$ (depending on the measurement method used and corrections for the
Galaxy and CG effect).  This anisotropy is at most 2 times larger than the
expected X-ray CG dipole. Applying Equation 19 the dipole measurements imply
that $1/7.1 \leq \Omega^{0.6}_0/b_x(0) \leq 1/1.7$. The favoured Method 2
dipole amplitude given in the last row of Table 2 is $\Delta=0.0065$ which
yields $\Omega^{0.6}_0/b_x(0)=1/4.8$.

The quantity $\Omega_{0}^{0.6}/b_x(0)$ has also been estimated from studies of
the X-ray selected AGN dipole under certain assumptions about local dynamics.
Generally $\Omega_{0}^{0.6}/b_x(0)$ ranges from $1/3.5$ if all the local
gravitational acceleration is assumed to arise from the volume with $R\ltsim
45$ h$^{-1}$Mpc, to $1/7$ if only half of the acceleration arises from within
this volume (\cite{miy94t}). Using the new IRAS-PSCz survey, Schmoldt et al
(1999) predict that some 65\% of the LG acceleration is generated within
40h$^{-1}$Mpc and that convergence is not reached until $\sim 140$h$^{-1}$Mpc.
Therefore $\Omega_{0}^{0.6}/b_x(0)$ is almost certainly larger than $1/3.5$
using this method. These results are in good agreement with our above
constraints from bulk flows and the HEAO dipole.  The observed HEAO dipole
therefore appears to be quite compatible with current measurements of the bulk
flow  of the local $\sim 60$h$^{-1}$Mpc volume. The dominant population of
X-ray emitters (AGN) in the 2-10 keV band is then highly biased
with respect to other tracers, e.g. optical or IRAS galaxies.

Over larger scales ($\sim 100-150$h$^{-1}$ Mpc) there is less agreement on the 
reality of bulk flow measurements. For example, the work of Lauer \& Postman
(1994) has suggested, with much controversy, that the Local Group has a motion
relative to the $z<0.05$ Abell cluster frame of 561$\pm 284$ kms$^{-1}$ in a
direction $l=220^{\circ}$, $b=-28^{\circ} (\pm 27^{\circ})$. Assuming a
dynamical origin of the observed CMB dipole this implies that the Abell cluster
frame (to $z=0.05$) is {\em itself} moving in bulk with respect to the CMB
frame with velociy of 689$\pm 178$ kms$^{-1}$ towards $l=343^{\circ}$,
$b=52^{\circ} (\pm 23^{\circ})$. If correct this could imply that $\sim 50$\%
of the Local Group motion is due to matter on scales $>100$h$^{-1}$Mpc. This
specific result has been strongly refuted by several other  works (e.g. Riess
et al 1995, Giovanelli et al 1997, Hudson et al 1999, Muller et al 1998, Dale
et al 1999). More recently however, independent observational evidence for 
bulk motion over these scales has emerged, in the range of $\sim 600-700$ km
s$^{-1}$ (e.g. Hudson et  al 1999, Willick 1998).  All such studies however
obtain directions for these motions greater than $60^{\circ}$  away from the LP
result, and are themselves highly fraught with potential systematic effects.

In criticism of these results it can be noted that there is an inconsistency
between (for example) the Lauer \& Postman measurement and the results of
gravity dipole estimation using galaxy catalogues. For example, the results of
Strauss et al (1992) using the 1.2Jy IRAS redshift survey found an
extraordinary convergence of the {\em direction} of the inferred gravity dipole
out to $\sim 20,000$kms$^{-1}$. This convergent dipole direction is only some
20$^{\circ}$ from the CMB dipole direction. (There are good arguments why the
velocity dipole of the Local Group is not necessarily converged until $z\sim 1$
(~\cite{pea92}), but that does not preclude a genuine convergence in a smaller
volume). Recently the IRAS-PSCz survey has largely confirmed these observations
(Schmoldt et al 1999). 

Scaling our above estimates for the HEAO XRB dipole anisotropy we predict that,
if all X-ray sources within $\sim 100-150$h$^{-1}$Mpc were removed then a 700
km s$^{-1}$ bulk flow would correspond to $0.0054 \leq \Delta \leq 0.0225 $
(assuming the measured range of allowed $\Omega_0^{0.6}/b_x(0)$). This would be
some $1.5-5.0$ times larger than the expected X-ray CG dipole amplitude.

The directions of both the bulk flow estimated from other works, and our
present dipole measurements are however scattered over a large area of the sky.
Figure 13 summarizes most of these directions. As already mentioned, the LP
flow direction is $\gtsim 60^{\circ}$ from all others, the Hudson et al (1999)
result is also significantly further from the Solar CMB velocity direction.
Interestingly the HEAO1-A2 measurements appear to be somewhat intermediate to
the LP result and the majority of the other, more local volume, estimates (SFI,
MIII, PSCz). However, recalling that we estimate an XRB dipole direction error
of at least $\sim 30^{\circ}$ for either method, then the HEAO dipole
directions are not inconsistent with (for example) the SFI and MIII flows.

 \section{Summary and Conclusions}

 The HEAO1-A2 X-ray data offers the best all-sky survey of baryonic matter to
$z\simeq 4$ currently available. Although  low level anisotropies in the X-ray
sky background arise largely from Galactic contributions, relatively crude
foreground removal clearly demonstrates the presence of extragalactic emission
associated with well known large scale structure (Virgo, Coma, Centaurus/Great
Attractor etc.) in the local universe. 

Qualitative comparison of the RASS 1.5 keV data with the 4 HEAO1-A2 bands used
here suggests that at least one third of the faint, unresolved HEAO1-A2
structure may be Galactic in origin, and possibly associated with the Galactic
Bulge.

The local extragalactic hard X-ray emission is dominated by AGN and galaxy
clusters. If we remove the flux associated with these sources to a flux limit
of $2 \times 10^{-11}$ erg s$^{-1}$ cm$^{-2}$ (2-10 keV) (which removes all
sources more luminous than $5.2 \times 10^{42}$ erg s$^{-1}$ (2-10 keV) out to
$z \simeq 0.015$) we measure a dipole anisotropy of $\Delta= 0.0023 $ to 0.0095
(depending on the method used and the details of the data processing). This
range of anisotropy is consistent with our expectations (LPT 97) of comparable
amplitude CG and LSS dipoles. It is significantly smaller than that measured in
the RASS 1.5 keV all-sky data by Plionis \& Georgantopoulos (1999). However, we
have argued that the hard ($>2$ keV) band XRB suffers less from Galactic
contamination and have shown how removal of the foreground of bright sources
reduces the dipole amplitude and shot noise (see also Treyer et al 1998).

We have derived the fully cosmological expressions for X-ray dipole anisotropy.
Unlike the often used Euclidean case, the relationship of the local
acceleration to the dipole anisotropy is no longer straightforward. However, in
the case of the current HEAO dataset we show that for  reasonable choice of
cosmology and matter density fluctuation power spectrum most of the dipole
anisotropy arises from $z\ltsim 0.1$ and the low-redshift, linear theory
approximation can be used.

Using current estimates of the bulk flow of the local 60h$^{-1}$ Mpc radius
volume and our XRB dipole measurements we find that $1/7.1 \leq
\Omega^{0.6}_0/b_x(0) \leq 1/1.7$. With our preferred dipole anisotropy
measurement then $\Omega^{0.6}_0/b_x(0)=1/4.8$. This implies that the
population of X-ray sources is highly biased with respect to optical or
infra-red selected objects. Studies of the dipole anisotropy of the local AGN
distribution (Miyaji 1994), and other analyses of the HEAO data (Boughn et al
1998, assuming epoch-independent biasing) also yield high values. Interestingly
our previous analysis of the angular power spectrum of the HEAO dataset (Treyer
et al 1998) which included terms as high as $l=20$ yielded a present-epoch
biasing factor $b_x(0)\sim 1-2$. The model fit to this data was however not
particularly good, and the lower order harmonics ($l=1-3$) are better fit with
a higher $b_x(0)$. If $\Omega_0 = 0.3$ then the values of $b_x(0)$ estimated
from the HEAO dipole/bulk flows fall into this lower range, however our
formalism is all based on an $\Omega=1$, Einstein-de-Sitter cosmology. We also
note that in all conventional models the bulk flow amplitude of a sphere with
radius $R$ drops with $R$ (specifically, if $P(k)\propto k^n$ then $V_{bulk}
\propto R^{-(n+1)/2}$). Therefore, if we overestimate the volume within which
we remove X-ray emission, but continue to apply the observed bulk flow
amplitudes for a larger sphere, we will then underestimate
$\Omega^{0.6}_0/b_x(0)$ from Equation 15.

If $\sim 700$ km s$^{-1}$ bulk flows over $100-150$ h$^{-1}$ Mpc radius volumes
did exist, as suggested by some studies (e.g. Lauer \& Postman 1994 , Willick
1998, Hudson et al 1999), then we predict that an XRB dipole anisotropy of
$0.0054 \leq \Delta \leq 0.0225 $ would be seen after removing source emission
within this volume.

It is worth noting that we should not discount further complications such as a
spatially varying local X-ray emmissivity to mass biasing.  Indeed, in their
study of the X-ray properties of the Great Attractor region and the Shapley
supercluster Raychaudhury et al (1991) found that for these similarly massive
regions the number counts of X-ray luminous clusters is quite different
(Shapley having the most). This is suggestive of a spatial variation in cluster
formation and brings into doubt the naive linear biasing scheme.

To fully exploit this, or future, hard X-ray all-sky data for cosmological or
dipole studies a better knowledge of the foreground contamination is essential.
In particular, a significantly more detailed model of the Galactic (and local,
e.g. LMC, SMC) emission is needed. Probably the best way to achieve this will
be through the use of softer band data (to provide spatial parameters) combined
with point-by-point spectroscopic information which will allow extrapolation
to the harder, less contaminated X-ray bands. A spectroscopy capable all-sky
imaging survey, such as that discussed by Jahoda (1998) would be well suited to
this (see also discussion in Treyer et al (1998)).

\acknowledgements This research has made use of data obtained through the High
Energy Astrophysics Science Archive Research Center (HEASARC), provided by the
NASA Goddard Space Flight Center. CAS acknowledges the support  of a National
Research Council fellowship, the University of Maryland, and NASA grant
NAG5-3257 over periods during which this work was undertaken. CAS and MT thank
the Racah Institute of Physics at the Hebrew University, Jerusalem, for its
hospitality during completion of this work.

 {}

\clearpage

 \begin{table}
  \begin{tabular}{lrll} Dataset & HEAO band & 
 Amplitude ($\Delta$) & Direction \\ \tableline Galactic
 Mask & Soft & 0.0132 & ($325^{\circ}, 51^{\circ}$) \\
      & Hard & 0.0084 & ($355^{\circ}, 45^{\circ}$) \\
      & Total& 0.0114 & ($327^{\circ}, 51^{\circ}$) \\
      & R15  & 0.0111 & ($324^{\circ}, 53^{\circ}$) \\ 
      &      &       &\\
      
 Complete 
 Mask& Soft & 0.0036 & ($343^{\circ}, 35^{\circ}$) \\
     & Hard & 0.0060 & ($29^{\circ} ,  2^{\circ}$) \\
     & Total& 0.0036 & ($345^{\circ}, 43^{\circ}$)  \\ 
     & R15  & 0.0030 & ($346^{\circ}, 48^{\circ}$)  \\ 
     &      &        & \\
 Complete 
 Mask& Soft & 0.0033 &  ($336^{\circ}, 40^{\circ}$) \\   
 3\% Iwan 
 model & Hard& 0.0032 & ($32^{\circ}, 2^{\circ}$) \\
 removed& Total& 0.0034 &($338^{\circ}, 47^{\circ}$) \\
        & R15 & 0.0029 & ($337^{\circ}, 54^{\circ}$) \\
        &      &       &  \\
 Complete 
 Mask      & Soft & 0.0030 & ($3^{\circ}, -3^{\circ}$) \\
 CG effect & Hard & 0.0048 & ($41^{\circ} ,  -27^{\circ}$) \\
 removed   & Total& 0.0027 & ($8^{\circ}, 5^{\circ}$)  \\ 
           & R15 & 0.0022 & ($15^{\circ}, 0^{\circ}$)  \\ 
          &     &    & \\
	  
  Complete 
 Mask      & Soft & 0.0025 & ($1^{\circ}, -4^{\circ}$) \\
 CG effect \& & Hard & 0.0045 & ($18^{\circ} ,  -44^{\circ}$) \\
 3\% Iwan   & Total& 0.0023 & ($6^{\circ}, 6^{\circ}$)  \\ 
 removed    & R15 & 0.0017 & ($15^{\circ}, 1^{\circ}$)  \\

 \end{tabular}  \caption{Spherical harmonic (Method 1) dipole
measurements of HEAO1-A2 data, directions in Galactic coordinates.}

 \end{table}
 
 \clearpage
 
  \begin{table}
 \begin{tabular}{lrl} Dataset & Amplitude
 ($\Delta$) & Direction \\ \tableline
 Galactic Mask & $0.0225 \pm 0.0005$& ($335^{\circ}\pm 24, 46^{\circ}\pm 12$) \\
      &   & \\
 Complete Mask & $0.0095 \pm 0.0005$& ($344^{\circ}\pm 24, 35^{\circ}\pm 12$) \\
      &    & \\ 
 Complete Mask & $0.0090 \pm 0.0005$& ($327^{\circ}\pm 24, 25^{\circ}\pm 12$) \\
 3\% Iwan model removed & & \\
     &  &   \\
 Complete Mask & $0.0085 \pm 0.0005$& ($353^{\circ}\pm 24, 1^{\circ}\pm 12$) \\
 CG effect removed &  &  \\
     &  &  \\
 Complete Mask & $0.0065 \pm0.0005$ & $(342^{\circ}\pm24, 7^{\circ}\pm 12$) \\
 CG effect \& & \\
 3\% Iwan removed & & \\
     
  \end{tabular} \caption{Least squares fit (Method 2)
  dipole amplitudes of Total band HEAO data. Errors correspond to finite search
 grid scales.} \end{table}
 
\clearpage

  \begin{table} 
  \begin{tabular}{lrll} Simulation &
 Dipole measure & Mean Amplitude ($\Delta$) & Mean sepn. from
 $(265^{\circ},48^{\circ}$)\\ \tableline 
 
 Complete Mask, & Method 1 & $0.00135 \pm 0.00031$ & -- \\ 
 randomized fluxes & Method 2 & $0.0022 \pm 0.0014 $ & -- \\
   &          &         \\
 + 4.2$\times 10^{-3}$ CG effect & Method 1 & $0.0024 \pm 0.0007$&
 $31^{\circ}\pm 20^{\circ}$ \\
   & Method 2 & $0.0043 \pm 0.0015$ & $25^{\circ}\pm 15^{\circ}$ \\
 \end{tabular} 
 
 \caption{Flux dipole measurements from simulations. Amplitudes are taken as
 the mean over 10 realizations, errors correspond to $1-\sigma$ standard
 deviations.}

 \end{table}

\clearpage

\begin{figure}
\protect\centerline{\psfig{figure=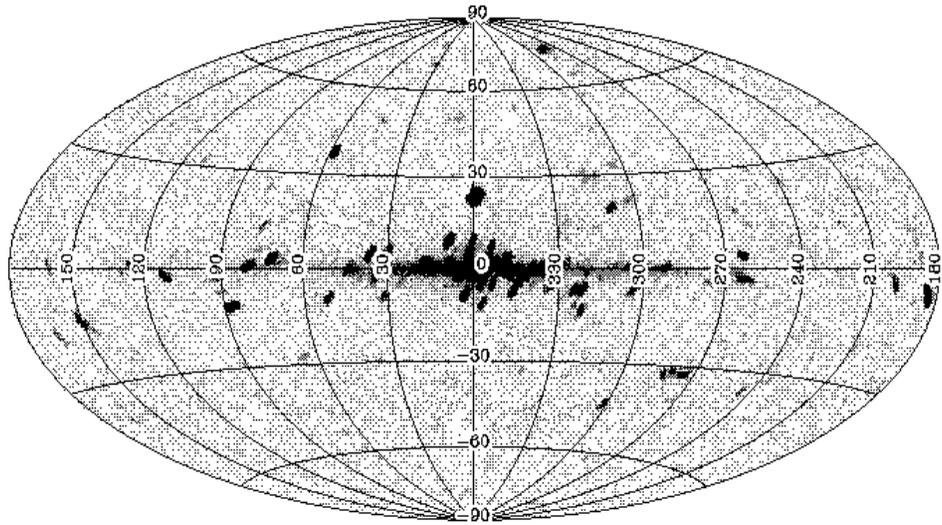,height=5.0truein,width=6.0truein,angle=-90}}

\caption{The full HEAO1-A2 Total band data shown in Hammer-Aitoff Galactic
coordinates projection. The Galactic Plane is clearly visible. Emission from
the Large and Small Magellanic Clouds is visible towards the bottom of the plot,
close to the southern ecliptic pole.}  \end{figure}

\begin{figure}
\protect\centerline{\psfig{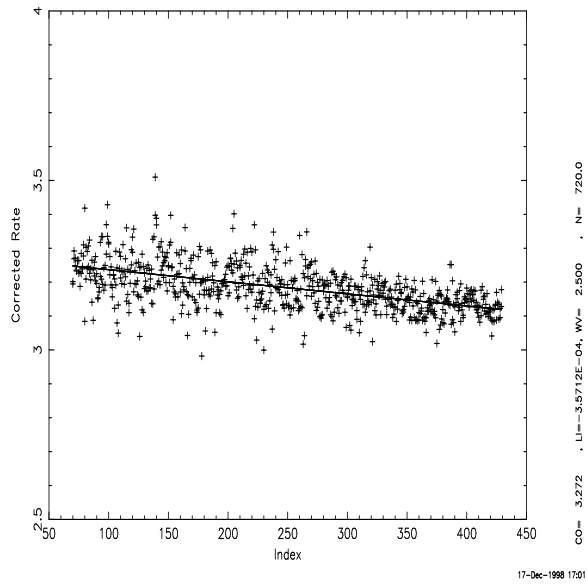}} 
\caption{The mean counts/sec averaged over the  time-wrapped Total band data
(subject to Galactic masking) as a function of longitudinal pixel index. The
least squares linear model is shown as the solid line.}
\end{figure}

\begin{figure}
\protect\centerline{\psfig{figure=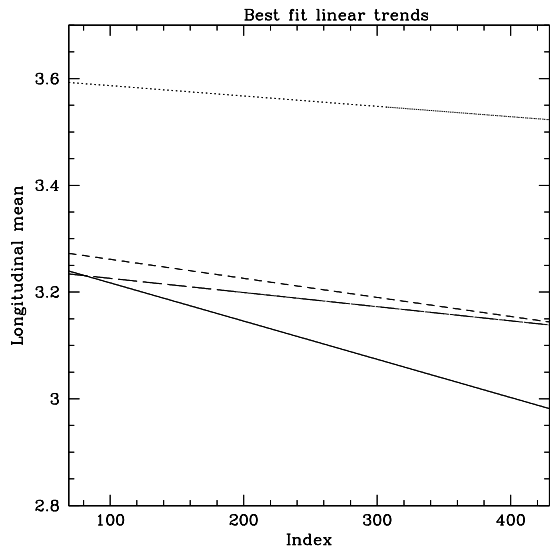,height=3.0truein,width=3.0truein,angle=0}}
\caption{The least squares linear trends to the time-wrapped HEAO data averaged
at each longitudinal index using the complete source and Galaxy mask. Solid
line - Soft band, dotted line - Hard band, short dashed line - Total band,
long-dashed line - R15 band.} 
\end{figure}
  
\begin{figure}
\protect\centerline{\psfig{figure=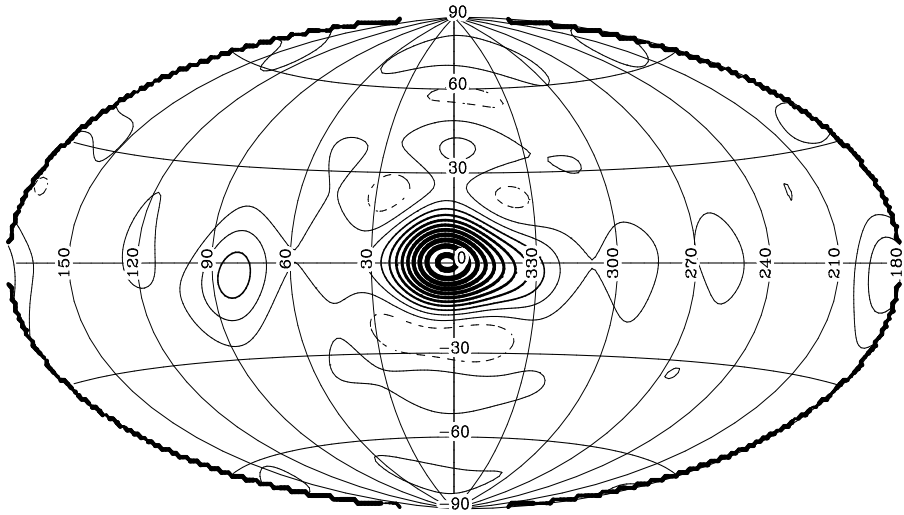,height=4.5truein,width=5.5truein,angle=-90}}

\caption{Harmonic reconstruction to $l_{max}=10$ (angular resolution $\sim
18^{\circ}$) of the raw total band HEAO1-A2 data in Galactic coordinates. Note
the complete domination by the Galaxy, in particular X-ray emission from the
region of the Galactic centre. The first solid contour is at the mean
intensity, subsequent soild contours are above the mean (dashed below) with
separation of $\sim 72$\% of the mean.  The uppermost solid contour is at
$\sim$ 720\% of the mean.}  
\end{figure}

\begin{figure}
\protect\centerline{\psfig{figure=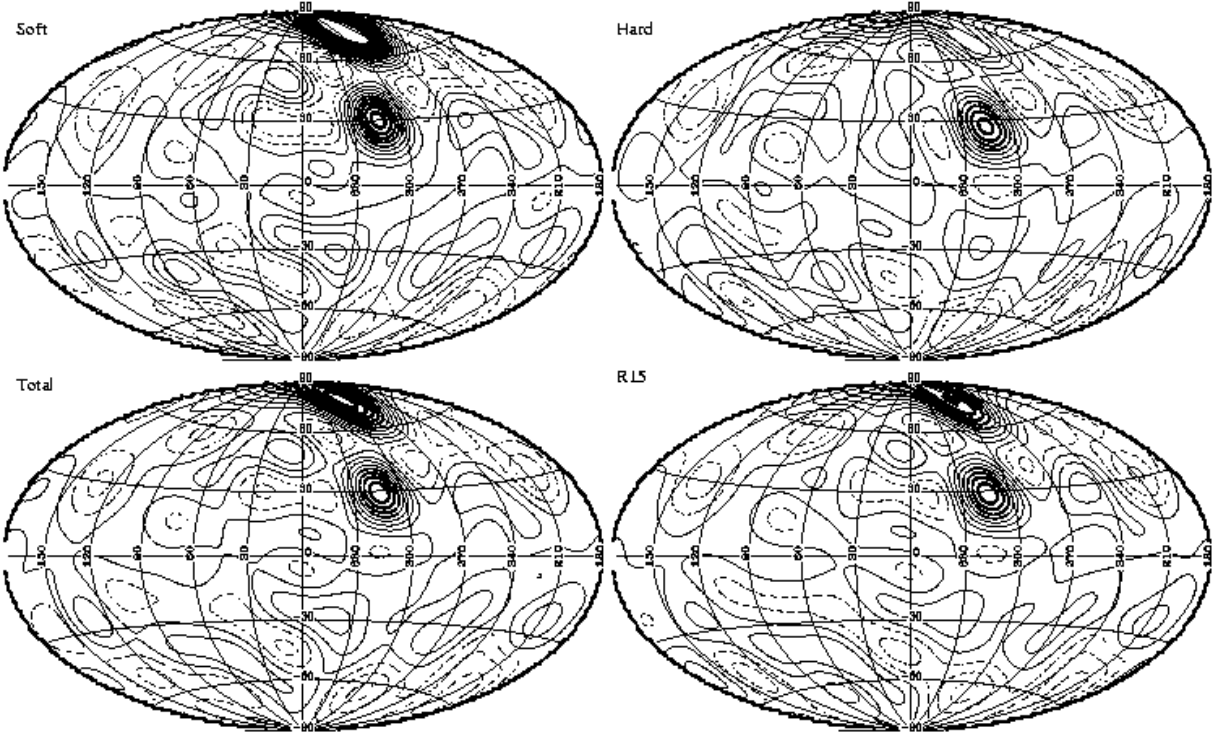,height=5.5truein,width=7truein,angle=-90}}

\caption{Harmonic reconstruction to $l_{max}=10$ in Galactic coordinates
(angular resolution $\sim 18^{\circ}$) of the 4 HEAO1-A2 bands with a Galactic
Plane cut of $|b|<22^{\circ}$ and all resolved Galactic sources removed as well
as the LMC nad SMC regions. Masked regions are filled at the mean intensity of
the unmasked regions. The first solid contour is at the mean intensity,
subsequent solid contours are above the mean intensity, dashed contours are
below the mean. Contour separation is 0.04 counts/s in all bands. Associated
structures can be identified: Virgo/Coma (at $b\gtsim 80^{\circ}$ and spanning
the Northern Galactic cap), Centaurus/Great Attractor (at $l\sim 315^{\circ}$
and $b\sim 30^{\circ}$)}
\end{figure}

\begin{figure}
\protect\centerline{\psfig{figure=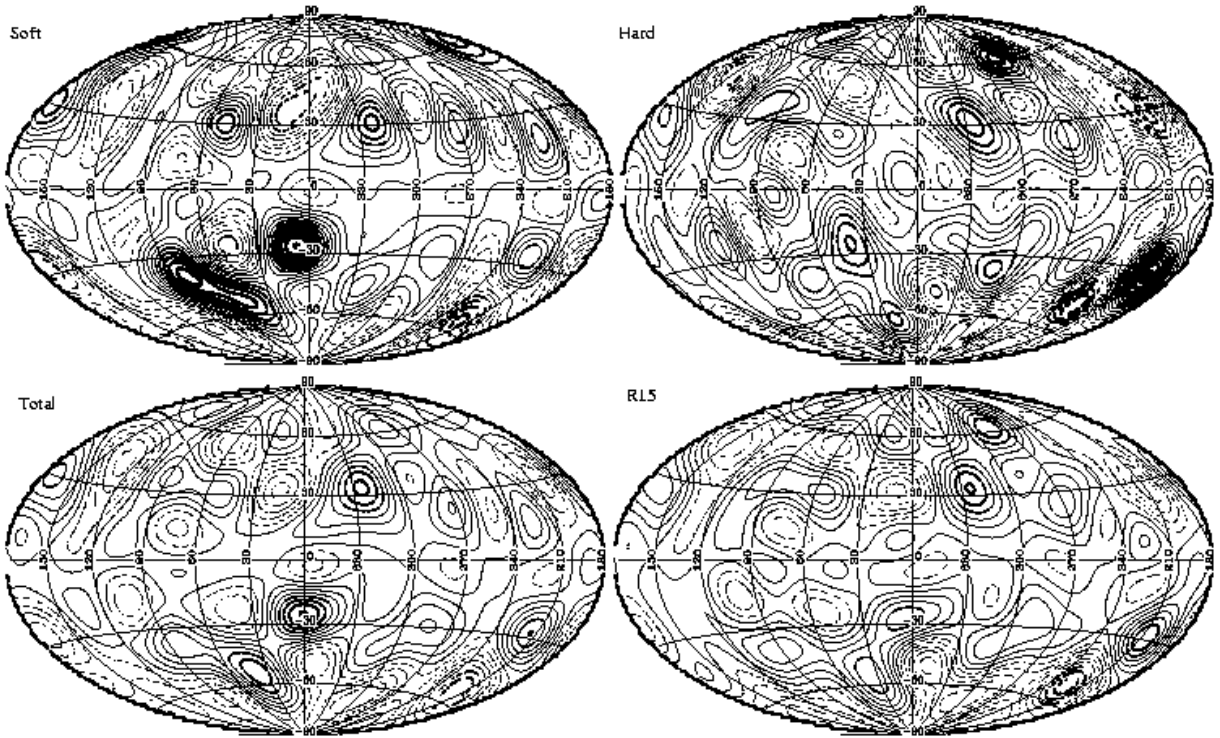,height=5.5truein,width=7truein,angle=-90}} 
\caption{Harmonic reconstruction to $l_{max}=10$ of HEAO data with the complete
mask applied. Contours are as for Figure 5 but now separated by 0.013
counts/sec. Some of the features seen may be associated with known structures.
For example, the feature at $l\sim 80^{\circ}, b\sim -45^{\circ}$ may be
associated wth the Pegasus cluster/supercluster which extends to $z\gtsim
0.04$, the soft feature at $l\sim 10^{\circ}, b\sim -30^{\circ}$ may be
Galactic (see text) although the Pavo-Indus-Telescopium structure lies in the
same direction. The feature at $l\sim 210^{\circ}, b\sim -30^{\circ}$ lies in
the direction of Abell 530 and 400, and the feature at $l\sim 280^{\circ},
b\sim 60^{\circ}$ matches the direction of the Leo cluster/group.} 
\end{figure}

\begin{figure}
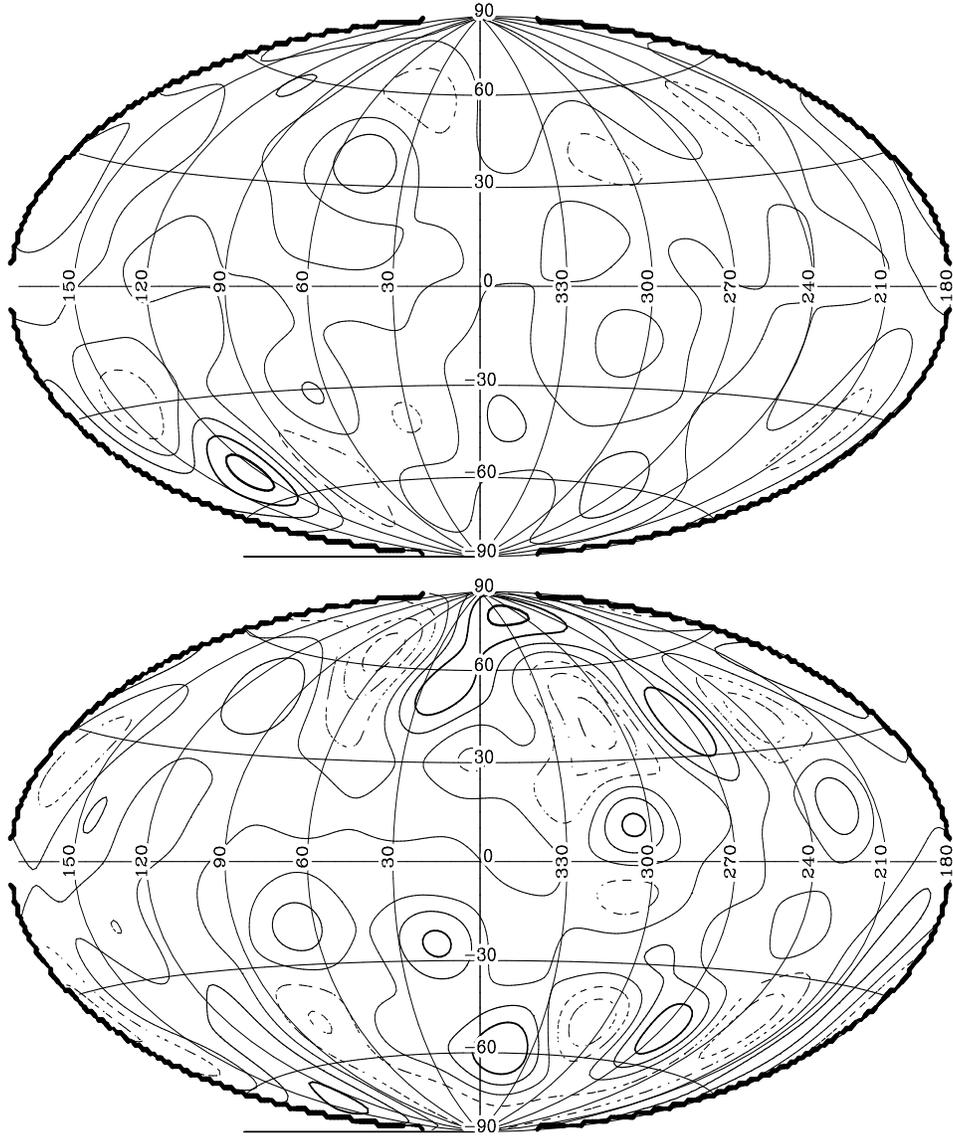

\protect\centerline{\psfig{figure=Figure_dip7a.ps,height=3truein,width=5truein,angle=-90}} 
\protect\centerline{\psfig{figure=Figure_dip7b.ps,height=3truein,width=5truein,angle=-90}} 

 \caption{In panels (a) and (b) the $l_{max}=10$ harmonic
reconstructions of  example total band `noise' skies are plotted for the two
mask cases in Figures 5 \& 6, using identical contour levels to these figures.
The spherical harmonic coefficients are generated from Monte-Carlo realisations
as described in Section 5 below, and reflect a random field drawn from the
appropriate angular power spectrum in both cases (Section 4). Sky masking is
also applied to these noise realisations.}

\end{figure}

\begin{figure}
\protect\centerline{\psfig{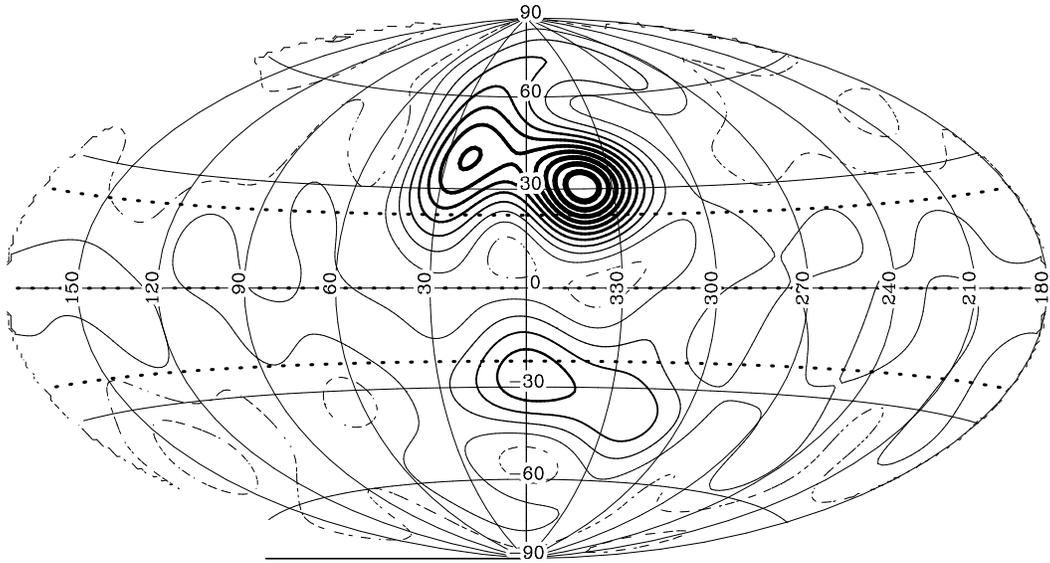}}

\caption{Harmonic reconstruction to $l_{max}=10$ of the RASS 1.5 keV data.
The complete HEAO1-A2 mask (Galactic and extragalactic sources removed) has
been applied to the data. Horizontal dotted lines at $b=\pm 20^{\circ}$ delimit
the Galactic plane mask. Contours are spaced at $\sim 1.7$\% of the monopole,
uppermost contour is at $\sim 17$\% of the monopole}
\end{figure}

\begin{figure}
\protect\centerline{\psfig{figure=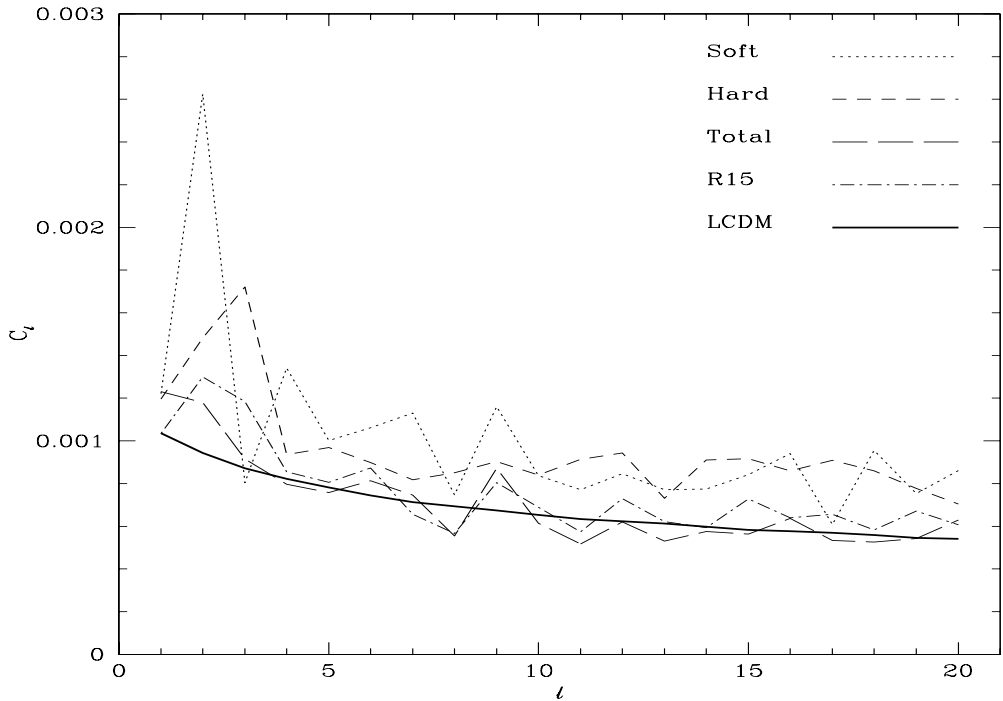,height=4.0truein,width=5.5truein,angle=0}}
\caption{The normalized spherical harmonic power spectrum of the HEAO1 XRB
is plotted to $l=20$ for Soft (dotted), Hard (short-dashed), Total
(long-dashed) and R15 (dot-dashed) colours. The Galactic mask and extragalactic
mask has been applied, corresponding to source removal to a flux limit of
$2\times 10^{-11}$ erg s$^{-1}$ cm$^{-2}$ (2-10 keV). The solid line shows a
fiducial 'best-fit' model for the Total band, as described in (\cite{tre98}): low
density CDM $P(k)$ with pure X-ray source density evolution and constant
biasing ($b_x(0)=1$), including the expected noise (see also \S 6).}
\end{figure}

\begin{figure}
\protect\centerline{\psfig{figure=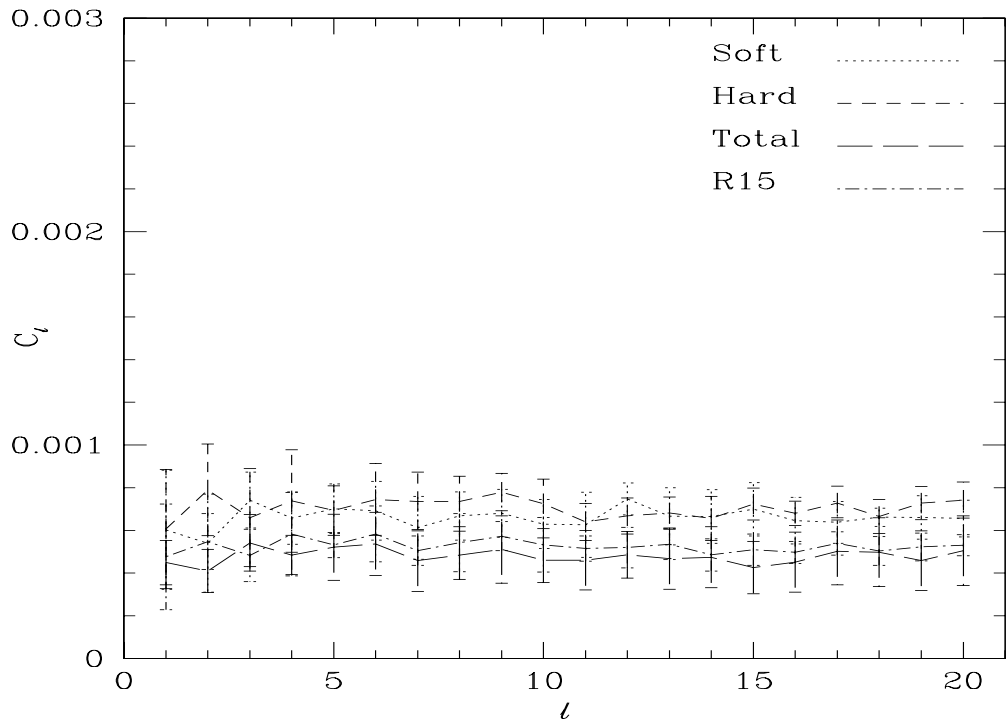,height=4.0truein,width=5.5truein,angle=0}}
\caption{The mean angular power spectrum over 10 realizations for each HEAO
colour is plotted for `noise' skys. Error bars correspond to the $1-\sigma$
scatter expected between individual realisations.  }
\end{figure} 

\begin{figure}  
\protect\centerline{\psfig{figure=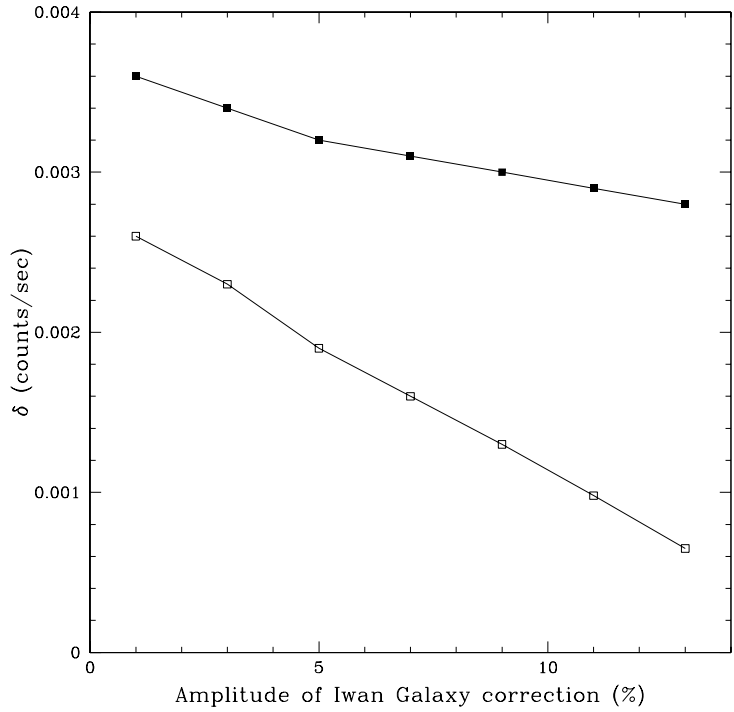,height=4.0truein,width=4.truein,angle=0}}
\caption{The Method 1 flux dipole amplitude in the Total band is plotted as
a function of increasing Galactic correction (Iwan model, normalised to \% of
monopole). The upper curve (solid squares) corresponds to the dipole without
removal of the expected CG effect, the lower curve (open squares) has had the
expected CG effect removed following the Iwan model Galactic correction.}
\end{figure}

\begin{figure}
\protect\centerline{\psfig{figure=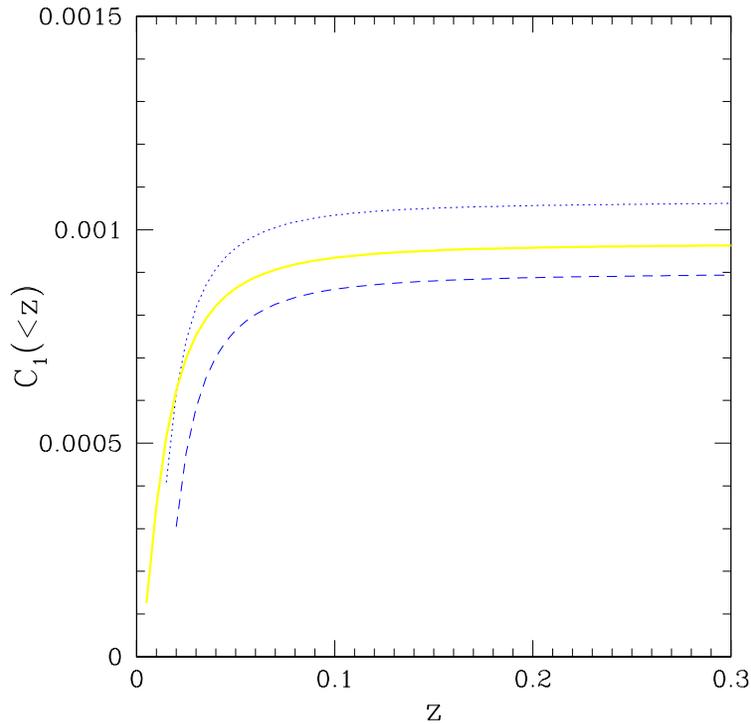,height=4.0truein,width=4.truein,angle=0}} 
\caption{The growth of the rms flux dipole is plotted as a function of outer
radius cut-off $z_{max}$ for a low density CDM $P(k)$ with pure density
evolution and constant biasing as described in (\cite{tre98}). The solid curve
assumes $z_{min}=0$ and a flux cutoff  $S_{cut}=2 \times 10^{-11}$ erg s$^{-1}$
cm$^{-2}$, while the thin lines simply assume a radial cutoff of $z_{min}=0.01$
(upper dotted line) and 0.015 (lower dashed line) respectively (for both cases 
$W_{cut}(z) \equiv 1$.} 
\end{figure}
 
\begin{figure}
\protect\centerline{\psfig{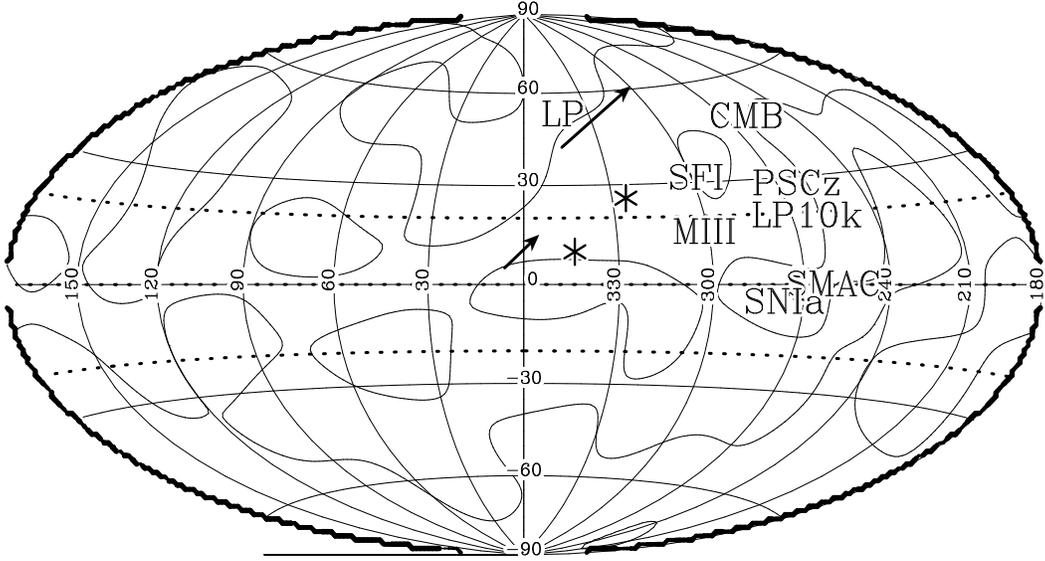}}
\caption{A summary of key directions and dipole results plotted in equal area
Galactic coordinates. Alphanumeric keys correspond to: LP - direction of Lauer
\& Postman (1994) bulk motion in CMB frame, CMB - direction of motion of the
Sun in CMB frame (direction of expected CG effect) (\cite{lin96}), SFI - bulk
flow direction of 50h$^{-1}$Mpc volume by Dale et al (1999) using SFI data
(Giovanelli et al 1997), PSCz - predicted bulk flow in same volume from IRAS
PSCz data (Schmoldt et al 1999), MIII - bulk flow in same volume from Mark III,
POTENT catalogue (Dekel et al 1999), SNIa - bulk flow in same volume from Riess
et al (1995), LP10k - bulk flow of 150 h$^{-1}$Mpc volume Tully-Fisher dataset
(Willick 1998), SMAC - bulk flow of 120h$^{-1}$Mpc volume (Hudson et al 1999).
Arrows correspond to Method 1 HEAO1-A2 dipoles with increasing Iwan Galactic
corrections (1-13\%), upper (higher latitude) and lower arrows correspond to
upper and lower curves in Figure 10 (without and with removal of the expected
CG effect, respectively). Upper and lower asterix's label the corresponding
Method 2 dipoles with 3\% Iwan model corrections (Table 2).} \end{figure}

 \end{document}